# Gesture Recognition for Feedback Based Mixed Reality and Robotic Fabrication: A Case Study of the UnLog Tower

**Alexander Htet Kyaw[1], Lawson Spencer[2], Sasa Zivkovic[2], and Leslie Lok[1]**

**Abstract.** Mixed Reality (MR) platforms enable users to interact with three-dimensional holographic instructions during the assembly and fabrication of highly custom and parametric architectural constructions without the necessity of two-dimensional drawings. Previous MR fabrication projects have primarily relied on digital menus and custom buttons as the interface for user interaction with the MR environment. Despite this approach being widely adopted, it is limited in its ability to allow for direct human interaction with physical objects to modify fabrication instructions within the MR environment. This research integrates user interactions with physical objects through real-time gesture recognition as input to modify, update or generate new digital information enabling reciprocal stimuli between the physical and the virtual environment. Consequently, the digital environment is generative of the user's provided interaction with physical objects to allow seamless feedback in the fabrication process. This research investigates gesture recognition for feedback-based MR workflows for robotic fabrication, human assembly, and quality control in the construction of the *UnLog Tower*.

[1]A. Kyaw, hk385@cornell.edu, L. Lok, wll36@cornell.edu (✉)
Rural-Urban Building Innovation (RUBI) Lab, Cornell University, Department of Architecture, Sibley Hall, Ithaca, NY 14850, United States

[2] L. Lawson, lls97@cornell.edu, S. Zivkovic, sz382@cornell.du
Robotic Construction Lab (RCL) Cornell University, Department of Architecture, Sibley Hall, Ithaca, NY 14850, United States

# 1 Introduction

Since the mid- 90s, Virtual Reality (VR) and Augmented Reality (AR) have existed under the umbrella of Mixed Reality (MR) on the Reality-Virtual (RV) Continuum between the absolute real environment and the absolute virtual environment (1). As VR and AR 3D user interfaces (3DUIs) have continued to become ubiquitous in architecture, construction, and academic research, the Milgram and Kashino's definition of MR has been further refined (2). In recent research, MR is often cited as an environment-aware overlay of digital content in the physical world where users are able to interact with both the digital and physical simultaneously (3). To facilitate human interaction between the digital and the physical MR environments, MR systems employ various techniques for collecting environmental and human physiological data, such as spatial mapping, hand-tracking, eye-tracking, and auditory recording. MR-enabled devices, such as the *Microsoft HoloLens 2* and *Meta Quest Pro*, utilize sensors, mics, and cameras to capture real-time data on changes in user behavior and the physical environment (4).

During the last decade, research using AR and MR workflows in the area of architectural fabrication have increased exponentially (5). Projects such as *Woven Steel*, *Timber De-Standardized, Code-Bothy*, and many more have explored human interaction with digital instructions in MR through digital interfaces such as buttons and menus or fiducial markers such as QR codes and AruCo markers (6–8). These MR fabrication projects have primarily focused on using human interactions with digital interfaces as the primary means to update the 3DUIs with new information. However, there exists an opportunity to directly incorporate human interaction with physical objects to update the 3DUI without needing digital interfaces.

This research integrates human interactions with physical objects through real-time gesture recognition as input to modify and update information in the digital environment. Through gesture recognition, touching a physical object could modify, update, or generate new digital information enabling seamless stimuli between the physical and the digital world. By recording user gestures as they interact with physical objects, the three-dimensional user interface can automatically provide new information in real time. As a result, the digital environment is generative of the user's provided interaction with physical objects. Through gestural tracking, user interactions with physical objects are recorded to determine the real-time location of physical objects in the digital environment. This can generate information such as localizing robotic tool paths, recognizing components, or measuring inaccuracies between the physical and the digital model. The real time generative data in the MR 3DUI allows the user to quickly respond to previous actions. The real time, feedback-based MR environment represents a cybernetic system whereby the outcome of interacting with a physical object(s) is taken as input for further action, thus creating a feedback loop until a desired condition is achieved.

The relationship between MR, gestural movement, digital twin, cybernetics, and human-computer interaction are used to help define systems of interaction between user and machine. From these relationships, the research presents three distinct *Gesture-Based Mixed Reality* (GBMR) fabrication workflows; a) *object localization* - registers the location of a physical object in the digital space, b) *object identification* - differentiates physical components using their digital parameters, c) *object calibration* - measures discrepancies between the physical object and associated digital geometry. Each of these three methods were used in six different tasks to construct the *UnLog Tower* (Fig 1). The workflows derivative of this research presents new opportunities for human-machine co-creation within physical and digital environments through MR in architecture and fabrication industries.

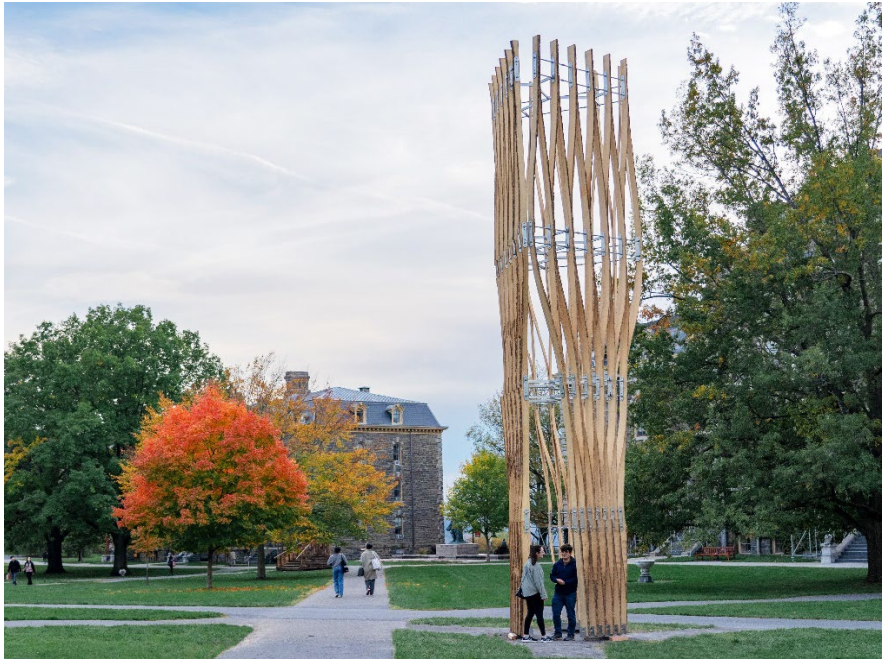

**Fig 1** The UnLog Tower, photo by Cynthia Kuo.

## 2 State of the Art

Innovative fabrication research projects such as *Holographic Construction*, *Code-Bothy*, *Woven Steel*, *Bent*, and *Timber De-Standardized 2.0*, use interactive "buttons" for users to toggle between different sets of digital geometry which is visible in the 3DUI (6–10). Though each of these projects use a *Microsoft Hololens* with *Fologram's* plug-in for *Rhino3d* and *Grasshopper*, the "buttons" can equally be interacted with one's mobile device. In each of these precedents, the "button" is a custom, pre-defined clickable digital object (either mesh or poly-surface). Thereby any change in the virtual interface is dependent on the user interacting with the select, pre-defined "buttons" or otherwise manipulating other digital geometry. *Holographic Construction* and *Code-Bothy* use digital "buttons" to toggle up and down between rows of bricks as they are laid (8,9). *Code-Bothy* has the added effect of color coordinating the amount of rotation per brick (8). *Woven Steel* and *Bent* exhibited several buttons to aid in the complex bending of tube steel and sheet metal (6,10). *Timber De-Standarized 2.0* developed menu list to visualize different aspects of an inventory of scanned irregular log meshes as well as cataloging and designing with the members through operations of slicing, indexing, baking, and isolating (7). Though these precedents offer an interaction between the user and the digital geometry, the interactions are limited to digital menus and buttons.

Other research projects such as *Augmented Feedback*, *Timber De-Standardized 1.0*, and *Augmented Vision* use various methods of AruCo markers for tracking, physics simulation, and real-time scanning to create an active responsive environment between digital and physical objects (11–13). In *Augmented Feedback,* AruCo makers were placed at nodal intersections of a bending-active bamboo grid-shell structure (11). AruCo marker tracking allowed users to digitize the locations of the markers and provide graphic feedback for all active users through the head mounted display (HMD). *Timber De-Standardized 1.0* utilized a physics simulation for fabricators to visualize and virtually "drop" irregular scanned meshes of logs till they found their resting point, which allowed for a precise alignment with its associated physical log (12). Finally, *Augmented Vision* uses the *Hololens 2* to track and scan the user's environment then display such information to inform the progress of constructing a minimal surface with strips of paper and/or bark (13). These projects have demonstrated the capabilities of feedback-based MR using additional systems such as AruCo markers, scanned meshes, and simulation.

Additionally, the accuracy of AR/MR platforms presents a significant challenge in many of these AR/MR fabrication workflows. The accuracy of the fabrication instructions provided to users depends on the precision of the system. As a result, several studies have been conducted to assess the accuracy of AR/MR systems. Researchers have investigated the use of AR for assembling metal pipes (14), weaving bamboo structures (15), and constructing complex wall systems with bricks within a tolerance of ± 20mm (16). Moreover, there have been research efforts aimed at improving the accuracy of AR/MR systems. A recent study by the authors explored the use of multiple QR codes to achieve a tolerance below 2mm with the Microsoft *HoloLens* 2 (17). The results of this study indicate that AR/MR systems have the potential to be used for high precision applications, such as assisting in robotic fabrication and accurate quality control.

## 3 Aim and Objectives

While previous MR projects have focused on using menus, AruCo markers, scanned meshes, and simulations to interact with digital geometries, this project investigates the potential of incorporating user's tactile interaction with physical objects as an input to update the 3DUI. Enabled by gesture recognition, this research demonstrates new methods to use both digital and physical stimuli for a generative MR fabrication experience. This research has developed 6 experiments to test 3 GBMR fabrication workflows for tasks such as generating geometry relative to physical objects, localizing robotic tool paths, recognizing discrete components according to parameters such as height and length, or measuring inaccuracies between the physical and the digital models. The methods for this research will first present the tools and software to conduct this research, which will then be followed by the three GBMR workflows used to fabricate the *UnLog Tower*: a) *object localization,* b) *object identification,* and c) *object calibration. Object localization* was used to determine the log geometry work object and the toolpath placement for robotic fabrication (Method 4.1). *Object identification* is utilized to identify physical components and display intuitive step-by-step assembly instructions (Method 4.2). *Object calibration* is employed to ensure the adjustment of jigs and the connection of panels match the digital model (Method 4.3). Each of these workflows will demonstrate new methods in MR research whereby physical stimuli can become a generative tool to interact and inform MR fabrication in real-time.

## 4 Methods

The following studies were conducted using Microsoft *HoloLens 2* and *Fologram*, a AR/MR plug-in for *Rhino3D* and *Grasshopper* (18–20). The near depth sensing camera on the Microsoft *HoloLens 2* is used for articulated hand tracking (AHAT). AHAT tracks the movement and gestures of the user's hand, independent from the visible light cameras (VLCs) used for simultaneous locating and mapping (SLAM). The articulated hand tracking system recognizes and records twenty-five 3D joint positions and rotations, including the wrist, metacarpals, proximal, distal, and fingertip joints (21). This data is live streamed from the *HoloLens 2* device to *Rhino3D* and *Grasshopper* via *Wi-Fi*. The Microsoft AHAT API provides access to the built-in gestural recognition algorithm of the *HoloLens 2*, enabling the utilization of its advanced capabilities for hand tracking purposes. The joint configuration and orientation obtained from AHAT can facilitate the estimation of hand poses, such as pinching, tapping, or poking (22). This study focuses on the use of pinching as the primary mode of gestural interaction by the user. The pinching gesture is recognized when the thumb tip and index fingertip are in close proximity (Fig 2). Additionally, a device capable of AHAT programming is imperative for gesture recognition and therefore is integral to the GBMR workflows.

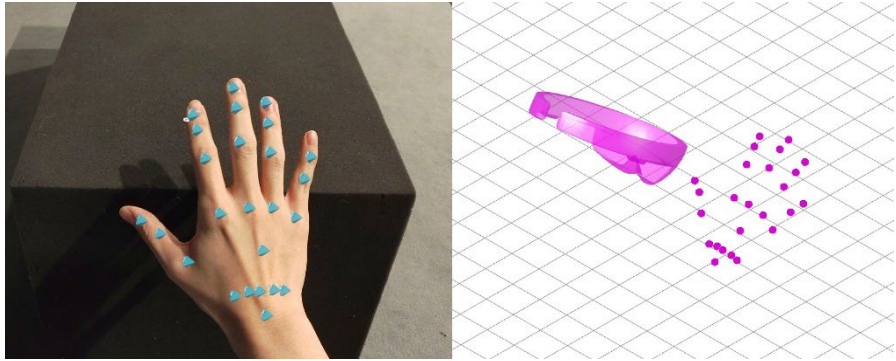

**Fig 2** Digital twin of *HoloLens 2* headset location, joint configuration, and orientation from AHAT (Articulated Hand Tracking); visualized through headset (left); visualized through *Rhino3D* and *Grasshopper* (right).

## *4.1 Object Localization*

The *UnLog Tower* exhibits robotically kerfed timber round woods that have been stretched along two threaded rods to form panels through a similar method exhibited at the *UnLog* pavilion at University of Virginia (23). Logs are irregular geometries that are comprised of knots and sometimes curved but can nonetheless be abstracted to a cylinder in most cases. Before the log is robotically kerfed, it is cut in half. To localize the robot targets to cut the log in half using a 6-axis robotic arm with a 5hp bandsaw end-effector, *object localization* method was employed. The user would place three points at both ends of the log to create two individual circles to generate a cylindrical mesh that was in line with the physical log (Fig. 3). Each point was created by the user pinching their right-hand index finger to their thumb. This feedback mechanism provides the user with a visual confirmation of the digitization process. From the cylindrical mesh, a surface was generated in the middle of the cylinder whereby the robot tool path could be derived from the robot targets at either end of the surface using *Robot Components* (24), a robot programming plug-in for ABB robots in *Grasshopper* that is then copied into *Robot Studio,* an ABB software for programming ABB robots (25).

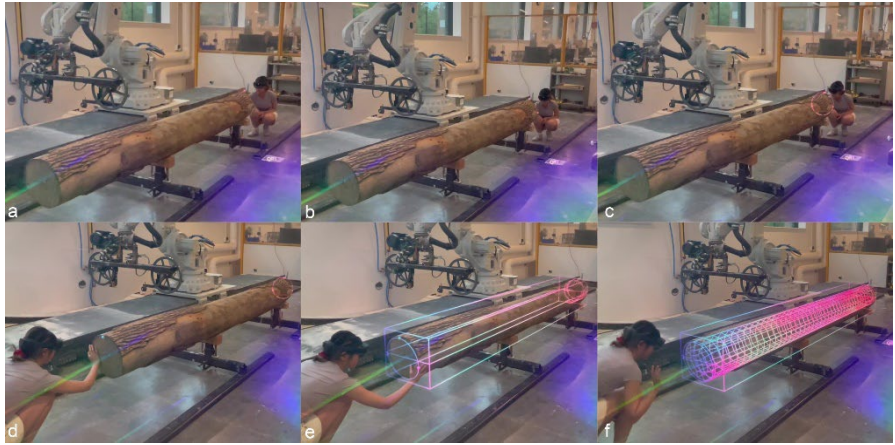

**Fig 3** *Object localization* is used to generate the location of a cylinder according to the diameter(s) of the log to automate the placement of the robotic toolpaths.

Once the log is cut in half, one half of the log is rotated 90° and remounted in the robot cell. According to the structural requirements for the *UnLog Tower* the cross section of each board was to be no less than 5" by 0.75". For each half log, the top and bottom ends of the log were to be trimmed off. The fabricator was to check the location of the cut surfaces within the log to ensure that the boards would meet the minimum cross-sectional requirements without any of the cut surfaces colliding with the 4" x 4" log mounts (Fig 4). Figure 4 demonstrates the process whereby the user can locate the half log in the robot cell by placing three points; two at one side of the half log to determine the diameter and one at the opposite end to determine the length of the half log. After the log geometry is defined, the user can set the location of the cut geometry by placing a point on the profile of the log (Fig 5). The MR system offers the user ongoing feedback during the process by performing a validation to determine whether the cut geometry falls within the boundary of the log. In the event that the cut geometry is placed outside the log or is situated too close to the log mount, a red notation with a cross mark is displayed within the 3DUI. The user may then respond to the alert and adjust the location of the cut geometry until a satisfactory outcome is achieved, represented by a green notation. The *object localization* workflow allows users to define points in the digital space that represents the physical log stock for work-object localization during robotic fabrication (Fig 6).

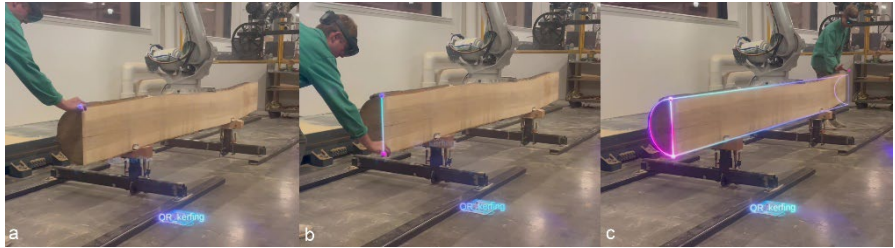

**Fig 4** *Object localization* is used to determine the work object placement for robotic fabrication.

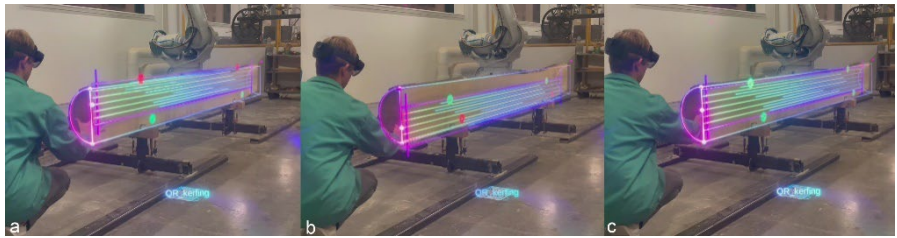

**Fig 5** *Object localization* is used to determine the placement of the toolpath for robotic fabrication.

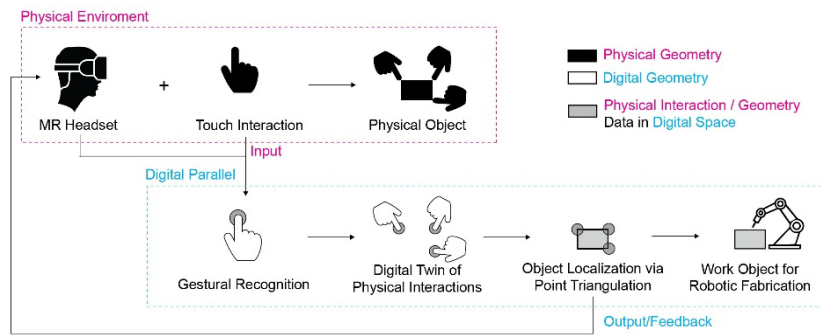

**Fig 6** *Object localization* system diagram describing how user interactions physical objects are used to create digital data through gestural recognition.

## *4.2 Object Identification*

*Object identification* is used to differentiate between self-similar physical components and display intuitive step-by-step assembly instructions. After the half logs have been robotically kerfed, they are set aside and prepared for finger jointing.

The finger joint template not only includes an outline for the finger joints, but also an outline for the hole that the threaded rod will ultimately pass through. Because of the parametric design of the kerfed timber panels for the *UnLog Tower,* the finger joint locations are staggard between adjacent boards within each half log. In order to correctly mark the location of the finger joints and the location for the threaded rod holes, GBMR was employed for *object identification* to correctly situate the location of the template per each board layer by registering the distance from the top of the board to the ground (determined by the QR code placement). The system determines the corresponding virtual template to display by comparing the calculated distance between the user-defined point to the ground with the predetermined distances of the virtual templates to the ground. The virtual template has an added notation that tells the user which layer they are on, so that the user can be sure that the physical template is being placed appropriately (Fig 7). The finger joints were cut with an oscillating saw and drill, while the holes for the threaded rods were drilled with a hole saw (Fig 8).

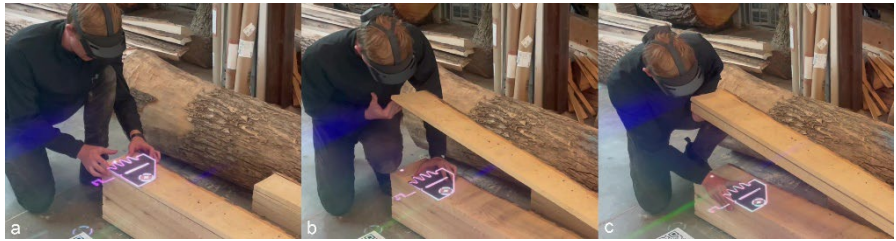

**Fig 7** *Object identification* is utilized to identify physical components and display intuitive step-by-step assembly instructions.

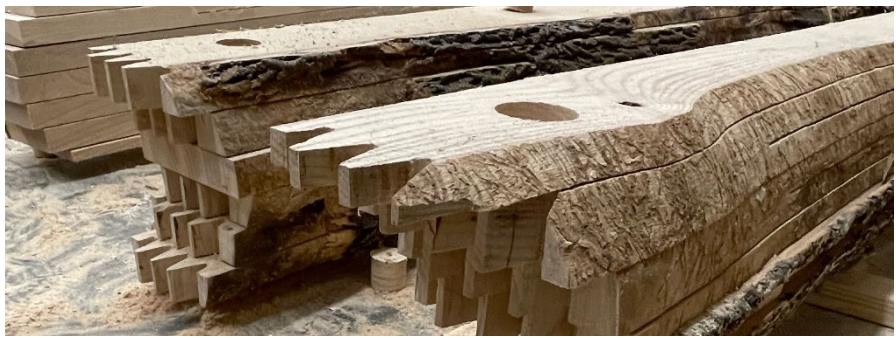

**Fig 8** Robotically Kerfed logs with finger joints and threaded rod holes

Additionally, *object identification* can be used to index and coordinate between self-similar parts. In order to brace the kerfed wood panels, the interior of the *UnLog Tower* exhibits 3 unique reciprocally steel tube frames. There are 9 unique tube lengths amongst 54 total steel tubes (Fig 9). After the steel tubes were cut to length, *object identification* was employed to index the tube steel according to their length and communicate to the user the location of the tube steel in the digital model(s) (Fig. 10). By placing a point at either end of the of the tube steel through gesture recognition, the user can define the length of the tube steel, which is checked against a list of tube steel lengths predetermined in the digital model. If the value between the user defined length and a predefined length is within a set tolerance of 0.5 inches, the 3DUI displays the corresponding digital information to the user through notation and two coordination models that visually indicate the location of the tube steel in the overall structure. The coordination model on the left (Fig 10b and 10c) illustrates at 1:1 scale the tube steel location within a particular tube steel frame and the coordination model on the right (Fig 10a, 10b, and 10c) illustrates at 1:10 scale a virtual model of the *UnLog Tower* with the location of the tube steel within the whole model. By using predetermined distances and gestural recognition, *Object Identification* can be used to pair digital assembly instructions with the identified physical object (Fig 11).

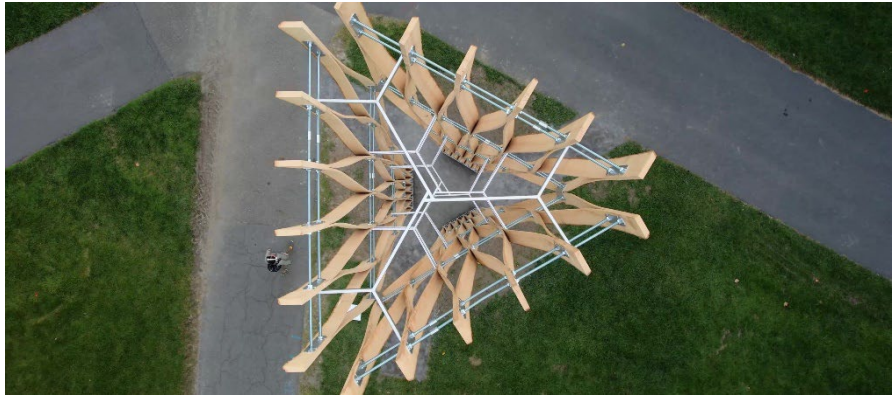

**Fig 9** Reciprocally framed tube steel in the *UnLog Tower,* photo by Cynthia Kuo.

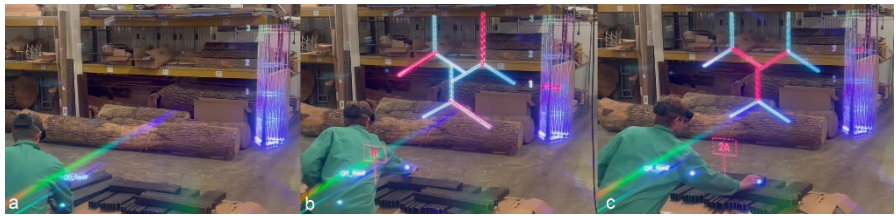

**Fig 10** *Object identification* is utilized to identify physical components and display part to whole assembly instructions.

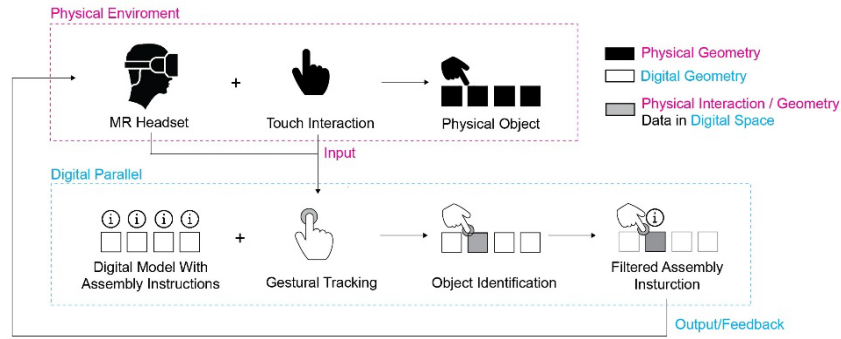

**Fig 11** *Object identification* system diagram describing how digital assembly is filtered through object identification via gestural recognition.

## *4.3 Object Calibration*

In order for the kerfed logs to splay out into panels, the threaded rods needed to have pre-located hex nuts appropriately placed to ensure that each board member would be in the correct location. In the GBMR workflow, *object calibration* was employed to place the hex nut locator correctly along a plywood jig. The hex nut locator was 3D printed with PLA to firmly hold each hex nut when it was screwed into the plywood board. A digital twin was created for each hex nut locator. When the user pinched the corner of the locator, *object calibration* would use gesture recognition to continuously track this movement, thereby synchronizing the digital geometry with the physical. As the physical object moved closer to the goal position, the notation would transform from red to yellow to green once the physical was properly located (Fig 12). This workflow represents a cybernetic system in which the adjustment of the physical locator position will generate new virtual feedback for the user, thus creating a feedback loop until the desired condition is attained. The desired condition is achieved when the digitized physical location of the hex nut locator is within a tolerance of 0.125 inches. This is indicated to the user via the notation system where the red or yellow cross turns into a green tick. The MR system will instruct the user to move onto the next hex nut locator only after the previous hex nut locator is correctly placed via gesture recognition. After all the hex nut locators were properly placed, a threaded rod is screwed through jig (Fig 13).

12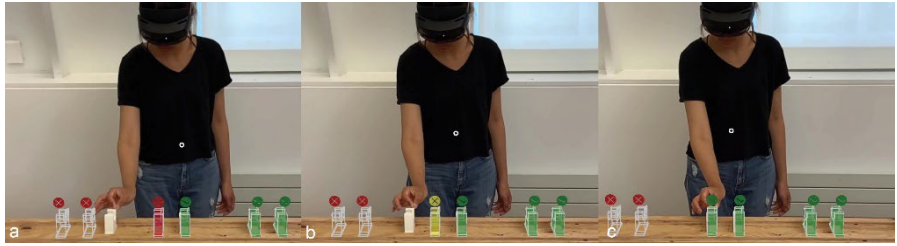

**Fig 12** *Object calibration* is employed to ensure the hex nut locators are adjusted to match the digital model. As the physical hex nut locator moves closer to its digital position, the notation would transform from red to yellow to green.

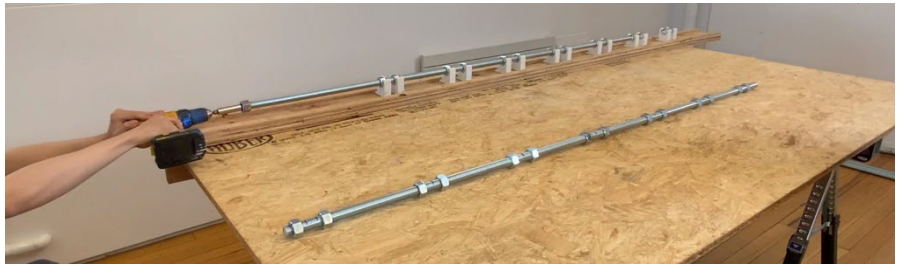

**Fig 13** After all the hex nut locators were properly placed, a threaded rod is screwed through jig

For the panel assembly, the robotically kerfed logs were splayed out along two threaded rods with pre-located hex nuts as was done in the *UnLog* pavilion (23). Temporary custom slip washers were placed between the hex nut and the board to ensure that the boards would keep their position until joined into larger prefab components with steel slip washers. Once panels were joined together in larger prefab components, *object calibration* was used to check the location of each board as they were tightened into location (Fig 14). This quality control step aligned a digital model of the goal geometry to the physical panel using the placement of a QR code. The physical location of the boards was determined by using GBMR to place a point at the center of the finger joint location of each board, which was automatically checked against the closest digital board. The deviation between the digitized board location and the digital board allowed for a 0.125" tolerance. A red cross notation indicates if the deviation was outside the tolerance, otherwise a green check notation would appear. This quality control step ensured that the parametrically defined wall panels were properly calibrated into larger prefab wall elements that were then transported to the site for assembly (Fig 15). By using the distance between physical and the digital object as variable, visual feedback is provided to the user during fabrication (Fig 16).

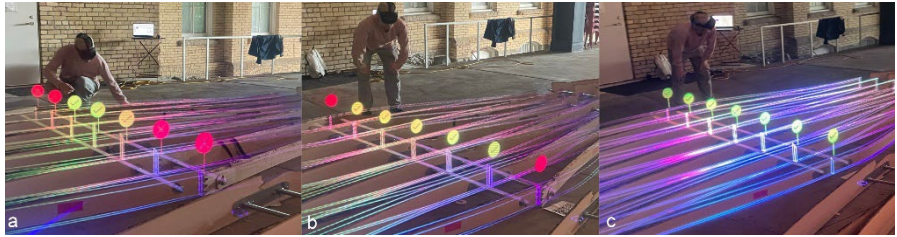

**Fig 14** *Object calibration* is employed for quality control of prefab wall components.

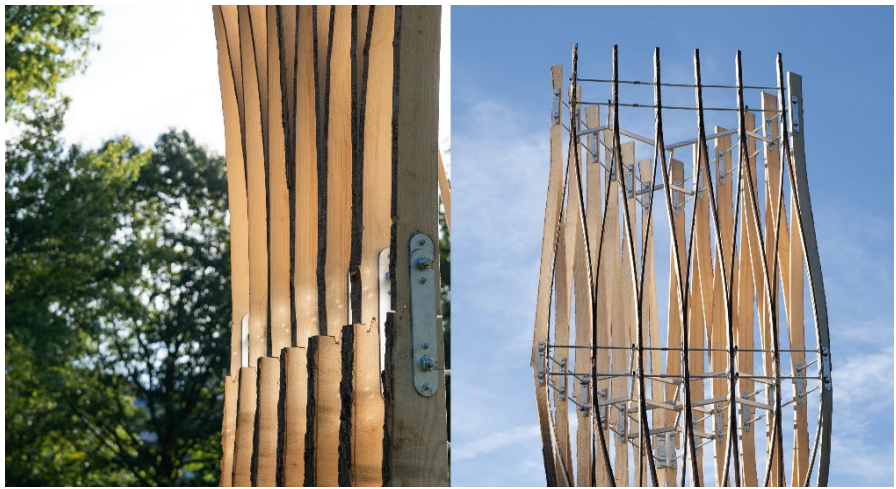

**Fig 15** Details of the *UnLog Tower*: finger joint splice connection (left) and robotically kerfed logs stretched along a thread rod (right), photos by Cynthia Kuo.

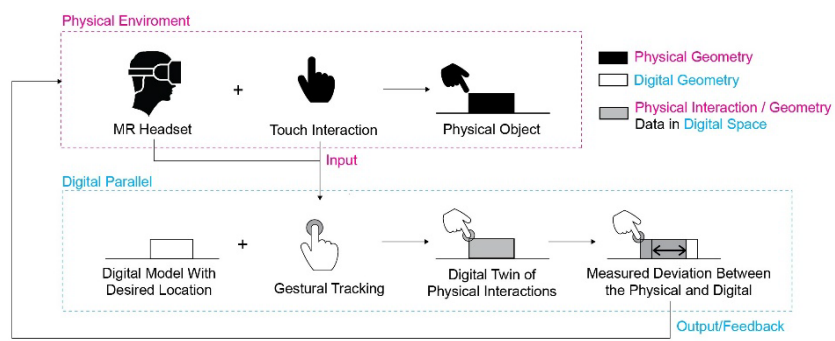

**Fig 16** *Object calibration* is employed to ensure the adjustment of jigs and the connection of panels match the digital model.

## 5 Results and Discussion

The implementation of gesture recognition for GBMR was incredibly useful for the fabrication of irregular and parametrically defined building components exhibited in the construction of the *UnLog Tower* (Fig 15). The prefab wall panels were attached to the tube steel reciprocal frames on site and lifted onto the foundation with a boom forklift.

The implementation of gesture recognition in MR fabrication workflows allowed users to define physical objects without the arduous placement of AruCo markers. The *object localization workflow* demonstrates that gesture recognition can be employed to locate robot work object data (Fig 6). However, the utilization of gesture recognition assumes a certain level of dexterity on the part of the user, as the data is dependent on the fidelity and accuracy of the user's fingers. The object localization workflow can be modified for robotic fabrication procedures that require a higher tolerance. Alternatively, improvements in the AHAT, articulated hand tracking, on the *Microsoft HoloLens 2* would also increase the accuracy of the overall system and the resolution of the work object placement.

The research also describes the potential of using gestural tracking for *object identification* whereby the user's hands can be intuitively used to index and coordinate objects between self-similar parts based upon predefined parameters (Fig 11). While this study utilizes the varying lengths of components as the parameter, future studies could begin incorporating the boundary geometry or volume in the workflow. This workflow holds enormous future potential for fabricators and programmers to develop future projects that employ this method to coordinate and educate subcontractors on the construction of parametric components with discretized or self-similar parts.

Finally, the *object calibration* workflow is a unique way for users to synchronize between physical objects and their digital twins (Fig 16). The threaded rod test is unique in that the user can pinch the hex nut locator while moving the physical object. Conversely, the second test with the panel quality control demonstrated that some objects are too heavy or cumbersome to pinch while moving. For that reason, the second test demonstrated the use of gesture recognition to iteratively define critical points until the physical geometry aligned with the digital model. As is the case with the *object localization* workflow, the accuracy of the gesture recognition is limited to the user's finger precision. This method will have to be modified for higher tolerance fabrication projects. Additionally, the method could have been employed to locate the foundation steel on the existing concrete slab that was used to support the project.

# 6 Conclusion

The future potential of using gesture recognition in MR fabrication projects is enormous. The presented research not only demonstrates that real time feedback through gesture recognition is imperative for advanced MR fabrication projects, but it can also be used in robotics, geometry creation, object indexing, model coordination, interactive digital twin, and complex quality control. Future investigations will seek to improve the accuracy of this method for high precision fabrication projects and explore the potential of incorporating a wider range of gestures, such as "tap", "poke", and "pinch". Additionally, a user-controlled interface is being developed to enable/disable or undo a recognized gesture.

The study highlights the potential of utilizing gestural recognition to innovate human-machine fabrication processes. Through real-time gesture tracking, GBMR workflows can seamlessly blend real and virtual environments with visual feedback and tactile interaction. The three GBMR workflows exhibited in this paper demonstrate the various applications for the real-time feedback-based fabrication and assembly of the *UnLog Tower*. This phygital experience offers a whole series of future applications investigations in the field of Mixed Reality fabrication and Human-Machine co-creation.